\documentclass{endm}
\usepackage{endmmacro}
\usepackage{graphicx}

\usepackage{diagbox}
\usepackage{makecell}
\newcolumntype{C}[1]{>{\centering\let\newline\\\arraybackslash\hspace{0pt}}m{#1}}

\usepackage{mathtools,booktabs}
\DeclarePairedDelimiter\floor{\lfloor}{\rfloor}
\usepackage{multirow}

\newcommand{\Nat}{{\mathbb N}}
\newcommand{\Real}{{\mathbb R}}

\begin{document}

\begin{verbatim}\end{verbatim}\vspace{2.5cm}

\begin{frontmatter}

\title{Computational determination of the largest lattice polytope diameter}

\author{Nathan Chadder
\thanksref{myemail}}
\address{Department of Computing and Software\\ McMaster University\\ Hamilton, Canada}

\author{Antoine Deza\thanksref{coemail}}
\address{Department of Computing and Software\\ McMaster University\\ Hamilton, Canada}
   \thanks[myemail]{Email:
   \href{mailto:chaddens@mcmaster.ca} {\texttt{\normalshape
   chaddens@mcmaster.ca}}} \thanks[coemail]{Email:
   \href{mailto:deza@mcmaster.ca} {\texttt{\normalshape
   deza@mcmaster.ca}}}

\begin{abstract}
A lattice $(d,k)$-polytope is the convex hull of a set of points in dimension $d$ whose coordinates are integers between $0$ and $k$. Let $\delta(d,k)$ be the largest diameter over all lattice $(d,k)$-polytopes. We develop a computational framework to determine $\delta(d,k)$ for small instances. We show that $\delta(3,4)=7$ and $\delta(3,5)=9$; that is, we verify for $(d,k)=(3,4)$ and $(3,5)$ the conjecture whereby $\delta(d,k)$ is at most $\lfloor(k+1)d/2\rfloor$ and is achieved, up to translation, by a Minkowski sum of lattice vectors.
\end{abstract}


\end{frontmatter}

\section{Introduction}\label{intro}
Finding a good bound on the maximal edge-diameter of a polytope in terms of its dimension and the number of its facets is not only a natural question of discrete geometry, but also historically closely connected with the theory of the simplex method, as the diameter is a lower bound for the number of pivots required in the worst case. Considering bounded polytopes whose vertices are rational-valued, we investigate a similar question where  the number of facets is replaced by the grid embedding size.

The convex hull of integer-valued points is called a lattice polytope and, if all the vertices are drawn from $\{0, 1, \dots, k\}^d$, it is referred to as a lattice $(d,k)$-polytope. Let $\delta(d,k)$ be the largest edge-diameter over all lattice $(d,k)$-polytopes. Naddef~\cite{Naddef1989} showed in 1989 that $\delta(d,1)=d$, Kleinschmidt and Onn~\cite{KleinschmidtOnn1992} generalized this result in 1992 showing that $\delta(d,k)\leq kd$. In 2016, Del Pia and Michini~\cite{DelPiaMichini2016} strengthened the upper bound to $\delta(d,k) \leq kd - \lceil d/2\rceil$ for $k\geq 2$, and showed that  $\delta(d,2)=\lfloor3d/2\rfloor$. Pursuing Del Pia and Michini's approach, Deza and Pournin~\cite{DezaPournin2016} showed that $\delta(d,k) \leq kd - \lceil 2d/3\rceil-(k-3)$ for $k\geq 3$, and that $\delta(4,3)=8$.  The determination of $\delta(2,k)$ was investigated independently in the early nineties by Thiele~\cite{Thiele1991}, Balog and B\'ar\'any~\cite{BalogBarany1991}, and Acketa and \v{Z}uni\'{c}~\cite{AcketaZunic1995}. Deza, Manoussakis, and Onn~\cite{DezaManoussakisOnn2017} showed that $\delta(d,k)\geq \lfloor (k+1)d/2\rfloor$ for all $k\leq 2d-1$ and proposed Conjecture~\ref{C3}.

\begin{conjecture}\label{C3}
$\delta(d,k)\leq \lfloor (k+1)d/2 \rfloor$, and $\delta(d,k)$ is achieved, up to translation, by a Minkowski sum of lattice vectors.
\end{conjecture} 

\noindent
In Section~\ref{CompFrame}, we propose a computational framework which drastically reduces the search space for lattice $(d,k)$-polytopes achieving a large diameter. Applying this framework to $(d,k)=(3,4)$ and $(3,5)$, we determine in Section~\ref{results} that $\delta(3,4)=7$ and $\delta(3,5)=9$.

\begin{theorem}
Conjecture~\ref{C3} holds for $(d,k)=(3,4)$ and $(3,5)$; that is, $\delta(3,4)=7$ and $\delta(3,5)=9$, and both diameters are achieved, up to translation, by a Minkowski sum of lattice vectors
\end{theorem}

\noindent
Note that  Conjecture~\ref{C3} holds for all known values of $\delta(d,k)$ given in Table~\ref{delta(d.k)}, and hypothesizes, in particular,  that $\delta(d,3)=2d$. The new entries corresponding to $(d,k)=(3,4)$ and $(3,5)$ are entered in bold.


\begin{table}[htb]
	\centering
	\begin{tabular}{|C{1cm}|C{0.75cm}C{0.75cm}C{0.75cm}C{0.75cm}C{0.75cm}C{0.75cm}C{0.75cm}C{0.75cm}C{0.75cm}C{0.75cm}|}
		\hline
		\theadfont\diagbox{$d$}{$k$}&
		1 & 2 & 3 & 4 & 5 & 6 & 7 & 8 & 9 & 10 \\
		\hline \hline
		1 & 1 & 1 & 1 & 1 & 1 & 1 & 1 & 1 & 1 & 1 \\
		2 & 2 & 3 & 4 & 4 & 5 & 6 & 6 & 7 & 8 & 8 \\
 		3 & 3 & 4 & 6 & {\bf 7}  & {\bf 9}  &   &   &   &   &   \\
		4 & 4 & 6 & 8 &   &   &   &   &   &   &   \\
		$\vdots$ & $\vdots$ & $\vdots$ &   &   &   &   &   &   &   & \\
		$d$ & $d$ & \multicolumn{1}{c}{$\floor{\frac{3d}{2}}$} &   &   &   &   &   &   &   & \\ 
		\hline
		\end{tabular}	
	\caption{The largest possible diameter $\delta(d,k)$ of a lattice $(d,k)$-polytope}\label{delta(d.k)}
\end{table}

\newpage
\section{Theoretical and Computational Framework}\label{CompFrame}
Since $\delta(2,k)$ and $\delta(d,2)$ are known, we consider in the remainder of the paper that $d\geq 3$ and $k\geq 3$.
While the number of lattice $(d,k)$-lattice polytopes is finite, a brute force search is typically intractable, even for small instances. Theorem~\ref{needed}, which recalls conditions established in~\cite{DezaPournin2016}, allows to drastically reduce the search space.

\begin{theorem}\label{needed}
For $d\geq 3$, let  $d(u,v)$ denote the distance between two vertices $u$ and $v$ in the edge-graph of a lattice $(d,k)$-polytope $P$ such that $d(u,v)=\delta(d,k)$.
For $i=1,\dots,d$, let $F_i^0$, respectively $F_i^k$, denote the intersection of $P$ with the facet of the cube $[0,k]^d$ corresponding to $x_i=0$, respectively $x_i=k$.
Then, $d(u,v)\leq \delta(d-1,k)+k$, and the following conditions are necessary for the inequality to hold with equality:
\begin{itemize}
\item[$(1)$] $u+v=(k,k,\dots,k)$,
\item[$(2)$] any edge of $P$ with $u$ or $v$ as vertex is $\{-1,0,1\}$-valued,
\item[$(3)$] for $i=1,\dots,d$, $F_i^0$, respectively $F_i^k$, is a $(d-1)$-dimensional face of $P$ with diameter $\delta(F_i^0)=\delta(d-1,k)$, respectively $\delta(F_i^k)=\delta(d-1,k)$.
\end{itemize}
\end{theorem}

\noindent
Thus, to show that $\delta(d,k) < \delta(d-1,k)+k$, it is enough to show that there is no lattice $(d,k)$-polytope admitting a pair of vertices $(u,v)$ such that $d(u,v)=\delta(d,k)$ and the conditions $(1)$, $(2)$, and $(3)$ are satisfied. The computational framework to determine, given $(d,k)$, whether $\delta(d,k)=\delta(d-1,k)+k$  is outlined below and illustrated for $(d,k)=(3,4)$ or $(3,5)$.

\newpage
\noindent
{\bf Algorithm to determine whether  $\delta(d,k) < \delta(d-1,k)+k$}\\\\
\noindent
{\em 
Step $1$: {\sc Initialization}\\
 Determine the set ${\cal F}$ of all the lattice $(d-1,k)$-polytopes $P$ such that $\delta(P)=\delta(d-1,k)$. For example, for $(d,k)=(3,4)$, the determination of all the $335$  lattice $(2,4)$-polygons $P$ such that $\delta(P)=4$ is straightforward.\\\\
Step $2$: {\sc Symmetries}\\
Consider, up to the symmetries of the cube $[0,k]^d$, the possible entries for a pair of vertices $(u,v)$ such that  $u+v=\{k,k,\dots,k\}$. For example, for $(d,k)=(3,4)$, the following 6 vertices cover all possibilities for $u$ up to symmetry: $(0,0,0), (0,0,1),(0,0,2), (0,1,1), (0,1,2)$, and $(0,2,2)$, where $v=(4,4,4) - u$.\\\\
Step $3$: {\sc Shelling}\\
For each of the possible pairs $(u,v)$ determined during Step $2$, consider all possible ways for $2d$ elements of the set ${\cal F}$ determined during Step 1 to form the $2d$ facets of $P$ lying on a facet of the  cube $[0,k]^d$. For example, for $(d,k)=(3,4)$ and $u=(0,0,0)$, we must find $6$ elements of ${\cal F}$,  $3$ with $(0,0)$ as a vertex, and $3$ with $(4,4)$ as a vertex. In addition, if an edge of an element of ${\cal F}$ with $u$ or $v$ as vertex is not $\{-1,0,1\}$-valued, this element is disregarded.\\\\\
Note that since the choice of an element of ${\cal F}$ defines the vertices of $P$ belonging to a facet of the cube $[0,k]^d$, the choice for the next element of ${\cal F}$ to form a shelling is significantly restricted. 
In addition, if the set of vertices and edges belonging to the current elements of ${\cal F}$ considered for a shelling includes a path from $u$ to $v$ of length at most $\delta(d-1,k)+k-1$, a shortcut between $u$ and $v$ exists and the last added elements of ${\cal F}$ can be disregarded.\\\\
Step $4$. {\sc Inner points}\\
For each choice of $2d$ elements of ${\cal F}$ forming a shelling obtained during Step $3$, consider the $\{1,2,\dots,k-1\}$-valued points not in the convex hull of the vertices of the $2d$ elements of ${\cal F}$ forming a shelling. Each such $\{1,2,\dots,k-1\}$-valued point is considered as a potential vertex of $P$ in a binary tree. If the current  set of edges includes a path from $u$ to $v$ of length at most $\delta(d-1,k)+k-1$, a shortcut between $u$ and $v$ exists and the corresponding node of the binary tree can be disregarded, and the the binary tree is pruned at this node.\\\\
A convex hull and diameter computation are performed for each node of the obtained binary tree. 
If there is a node yielding a diameter of $\delta(d-1,k)+k$ we can conclude that $\delta(d,k)=\delta(d-1,k)+k$. Otherwise, we can conclude that $\delta(d,k)<\delta(d-1,k)+k$.
For example, for $(d,k)=(3,5)$, no choice of $6$ elements of ${\cal F}$ forming a shelling such that $d(u,v)\geq  10$ exist, and thus Step 4 is not executed.
}

\section{Computational Results}\label{results}
For $(d,k)=(3,4)$, a shelling exists for which path lengths are not decidable by the algorithm without convex hull computations. However, this shelling only achieves a diameter of 7. For $(d,k)=(3,5)$  the algorithm stops at Step $3$, as there is no combination of $6$ elements of ${\cal F}$ which form a shelling such that $d(u,v)\geq  \delta(2,5)+5$. Thus, no convex hull computations are required for $(d,k)=(3,5)$. A shortcut from $u$ to $v$ is typically found early on in the shelling, which leads to the algorithm terminating quickly. Run on a 2009 Intel\textsuperscript{\textregistered} Core\texttrademark 2 Duo 2.20GHz CPU, the algorithm is able to terminate for $(d,k) = (3,4)$ and $(3,5)$ in under a minute. Consequently, $\delta(3,4)<8$ and $\delta(3,5)<10$. 
Since the Minkowski sum of $(1,0,0),(0,1,0),(0,0,1),(0,1,1),(1,0,1),(1,1,0)$, and $(1,1,1)$ forms a lattice $(3,4)$-polytope with diameter $7$, we conclude that $\delta(3,4)=7$. Similarly, since the Minkowski sum of $(1,0,0),(0,1,0),(0,0,1),$ $(0,1,1),(1,0,1),(1,1,0), (0,1,-1),(1,0,-1)$, and $(1,-1,0)$ forms, up to translation, a lattice $(3,5)$-polytope with diameter $9$, we conclude that $\delta(3,5)=9$. Computations for additional values of $\delta(d,k)$ are currently underway. In particular, the same algorithm may determine whether $\delta(d,k)=\delta(d-1,k)+k$ or $\delta(d-1,k)+k-1$ for $(d,k)=(5,3)$ and $(4,4)$ provided the set of all lattice $(d-1,k)$-polytopes achieving $\delta(d-1,k)$ is determined for  $(d,k)=(5,3)$ and $(4,4)$. Similarly, the algorithm could be adapted to determine whether $\delta(d,k)<\delta(d-1,k)+k-1$ provided the set of all lattice $(d-1,k)$-polytopes achieving $\delta(d-1,k)$ or $\delta(d-1,k)-1$ is determined. For example, the adapted algorithm may determine whether $\delta(3,6)=10$.

\begin{ack}
This work was partially supported by the Natural Sciences and Engineering Research Council of Canada Discovery Grant program (RGPIN-2015-06163).
\end{ack}


\newpage
\begin{thebibliography}{10}\label{bibliography}

\bibitem{AcketaZunic1995} 
	Dragan Acketa and Jovi\u{s}a \u{Z}uni\'c,
    	\emph{On the maximal number of edges of convex digital polygons included into an $m\times{m}$-grid}, 
    	Journal of Combinatorial Theory A \textbf{69} (1995), 358--368.


\bibitem{BalogBarany1991} 
	Antal Balog and Imre B\'ar\'any,
    	\emph{On the convex hull of the integer points in a disc}, 
	Proceedings of the Seventh Annual Symposium on Computational Geometry (1991), 162--165.



\bibitem{DelPiaMichini2016} 
	Alberto Del Pia and Carla Michini,
    	\emph{On the diameter of lattice polytopes},
     	Discrete and Computational Geometry \textbf{55} (2016), 681--687.

\bibitem{DezaManoussakisOnn2017} 
	Antoine Deza, George Manoussakis, and Shmuel Onn,
    	\emph{Primitive zonotopes}, 
    	Discrete and Computational Geometry (to appear).
	
\bibitem{DezaPournin2016} 
	Antoine Deza and Lionel Pournin, 
	\emph{Improved bounds on the diameter of lattice polytopes},
	arXiv:1610.00341 (2016).


\bibitem{KleinschmidtOnn1992} 
	Peter Kleinschmidt and Shmuel Onn, 
	\emph{On the diameter of convex polytopes}, 
	Discrete Mathematics \textbf{102} (1992), 75--77.

\bibitem{Naddef1989} 
	Dennis Naddef, 
	\emph{The Hirsch conjecture is true for $(0,1)$-polytopes}, 
	Mathematical Programming \textbf{45} (1989), 109--110.



\bibitem{Thiele1991} 
	Torsten Thiele, 
	\emph{Extremalprobleme f\"{u}r Punktmengen},
    	Master thesis, Freie Universit\"{a}t, Berlin, 1991.


    
\end{thebibliography}
\end{document}